\documentclass[8.5pt,twoside,twocolumn]{article}
\usepackage[super,sort&compress,comma]{natbib} 
\usepackage[version=3]{mhchem}
\usepackage{balance}
\usepackage{times,mathptm}
\usepackage{sectsty}
\usepackage{graphicx} 
\usepackage{lastpage}
\usepackage[format=plain,justification=raggedright,singlelinecheck=false,font=small,labelfont=bf,labelsep=space]{caption} 
\usepackage{fancyhdr}
\usepackage{amsmath}
\usepackage[table]{xcolor}
\usepackage{amsfonts,amssymb,latexsym}
\usepackage{bm}
\usepackage[mathcal]{euscript}
\usepackage{graphicx}
\usepackage{epsfig}
\usepackage{color}
\usepackage{booktabs}

\newcommand{\etal}{\textit{et al. }}

\newcommand{\leg}[1]{\textbf{#1}}

\newcommand{\sls}{\textit{saltless}}
\newcommand{\sout}{\textit{salt-out}}
\newcommand{\stin}{\textit{salt-in}}
\newcommand{\DDP}{D_{\rm DP}}
\newcommand{\DF}{D_{\rm eff}}

\begin{document}

\thispagestyle{plain}
\fancypagestyle{plain}{
\renewcommand{\headrulewidth}{1pt}}
\renewcommand{\thefootnote}{\fnsymbol{footnote}}
\renewcommand\footnoterule{\vspace*{1pt}%
\hrule width 3.4in height 0.4pt \vspace*{5pt}} 
\setcounter{secnumdepth}{5}

\makeatletter 
\renewcommand\@biblabel[1]{#1}            
\renewcommand\@makefntext[1]%
{\noindent\makebox[0pt][r]{\@thefnmark\,}#1}
\makeatother 
\renewcommand{\figurename}{\small{Fig.}~}
\sectionfont{\large}
\subsectionfont{\normalsize} 

\fancyfoot{}
\fancyfoot[RO]{\footnotesize{\sffamily{1--\pageref{LastPage} ~\textbar  \hspace{2pt}\thepage}}}
\fancyfoot[LE]{\footnotesize{\sffamily{\thepage~\textbar\hspace{3.45cm} 1--\pageref{LastPage}}}}
\fancyhead{}
\renewcommand{\headrulewidth}{1pt} 
\renewcommand{\footrulewidth}{1pt}
\setlength{\arrayrulewidth}{1pt}
\setlength{\columnsep}{6.5mm}
\setlength\bibsep{1pt}

\twocolumn[
  \begin{@twocolumnfalse}
\noindent\LARGE{\textbf{How a ``pinch of salt'' can tune chaotic mixing of colloidal suspensions$^\dag$}}
\vspace{0.6cm}

\noindent\large{\textbf{Julien Deseigne,\textit{$^{a}$} C\'ecile Cottin-Bizonne,\textit{$^{a}$} Abraham D. Stroock,\textit{$^{b}$} Lyd\'eric Bocquet,\textit{$^{a,c}$} and
Christophe Ybert~\textit{$^{a,\ddag,*}$}}}\vspace{0.5cm}

\noindent\textit{\small{\textbf{Received Xth XXXXXXXXXX 20XX, Accepted Xth XXXXXXXXX 20XX\newline
First published on the web Xth XXXXXXXXXX 200X}}}

\noindent \textbf{\small{DOI: 10.1039/b000000x}}
 \end{@twocolumnfalse} \vspace{0.6cm}

  ]

\noindent\textbf{Efficient mixing of colloids, particles or molecules is a central issue in many processes. It results from the complex interplay between flow deformations and molecular diffusion, which is generally assumed to control the homogenization processes. In this work we demonstrate on the contrary that despite fixed flow and self-diffusion conditions, the chaotic mixing of colloidal suspensions can be either boosted or inhibited by the sole addition of trace amount of salt as a co-mixing species. Indeed, this shows that local saline gradients can trigger a chemically-driven transport phenomenon, diffusiophoresis, which controls the rate and direction of molecular transport far more efficiently than usual Brownian diffusion. A simple model combining the elementary ingredients of chaotic mixing with diffusiophoretic transport of the colloids allows to rationalize our observations and highlights how small-scale out-of-equilibrium transport bridges to mixing at much larger scales in a very effective way. Considering chaotic mixing as a prototypal building block for turbulent mixing, this suggests that these phenomena, occurring whenever the chemical environment is inhomogeneous, might bring interesting perspective from micro-systems up to large-scale situations, with examples ranging from ecosystems to industrial contexts.}
\section*{}
\vspace{-1cm}
\footnotetext{\dag~Electronic Supplementary Information (ESI) available: Additional details on experimental and high resolution cross-section images of the mixing process for different salt configurations. See DOI: 10.1039/b000000x/}


\footnotetext{\textit{$^{a}$~Institut Lumi{\`e}re Mati\`ere, Universit\'e Claude Bernard Lyon 1-CNRS, UMR 5306, Universit\'e de Lyon, F-69622 Villeurbanne, France.}}
\footnotetext{\textit{$^{b}$~School of Chemical and Biomolecular Engineering, Cornell University, Ithaca, New York 14853, USA.}}
\footnotetext{\textit{$^{c}$~Present address: Department of Civil and Environmental Engineering, Massachusetts Institute of Technology, Cambridge, MA, USA}}
\footnotetext{\textit{$^{\ddag}$~E-mail for correspondence: Christophe.Ybert@univ-lyon1.fr}}


Transport and mixing of molecules or particles play a central role in many processes, from large scales phenomena such as industrial chemical reactors, dispersion of pollutants or deposition of sediments involved in bio\-chemical cycles \cite{dai1995signifi,lee1992organic}, down to the rapidly developing Lab-on-a-Chip micro-systems. In this field, strong limitation to mixing involved by small-scales viscous flows have triggered research for specific strategies \cite{Bringer2004, Schonfeld2004}  to optimize and facilitate (bio-)chemical analysis \cite{Burns1996} or synthesis \cite{Watts2003} operations in microsystems. The generic route for efficient mixing proceeds through the stretch and fold mechanism that generates ever-thinner structures of the substances to mix. This is achieved either from turbulent flow properties \cite{ottino1990mixing,villermaux2006coarse} in large-scales situations, or through laminar chaotic flows in viscous regimes \cite{stroock2002chaotic,raynal2007towards} and, accordingly, mixing has been mostly approached from the ``large'' hydrodynamic scales perspective, focusing on flow characteristics.
In all such processes however, mixing or homogenization involves a continuous interplay of global deformation within the flow, and local molecular transport --generically assumed to occur through Brownian diffusion \cite{villermaux2000mixing,villermaux2006coarse,stroock2002chaotic,Sundararajan2012}.
%

\begin{figure}[!t] 
\centering \includegraphics[width=0.47\textwidth]{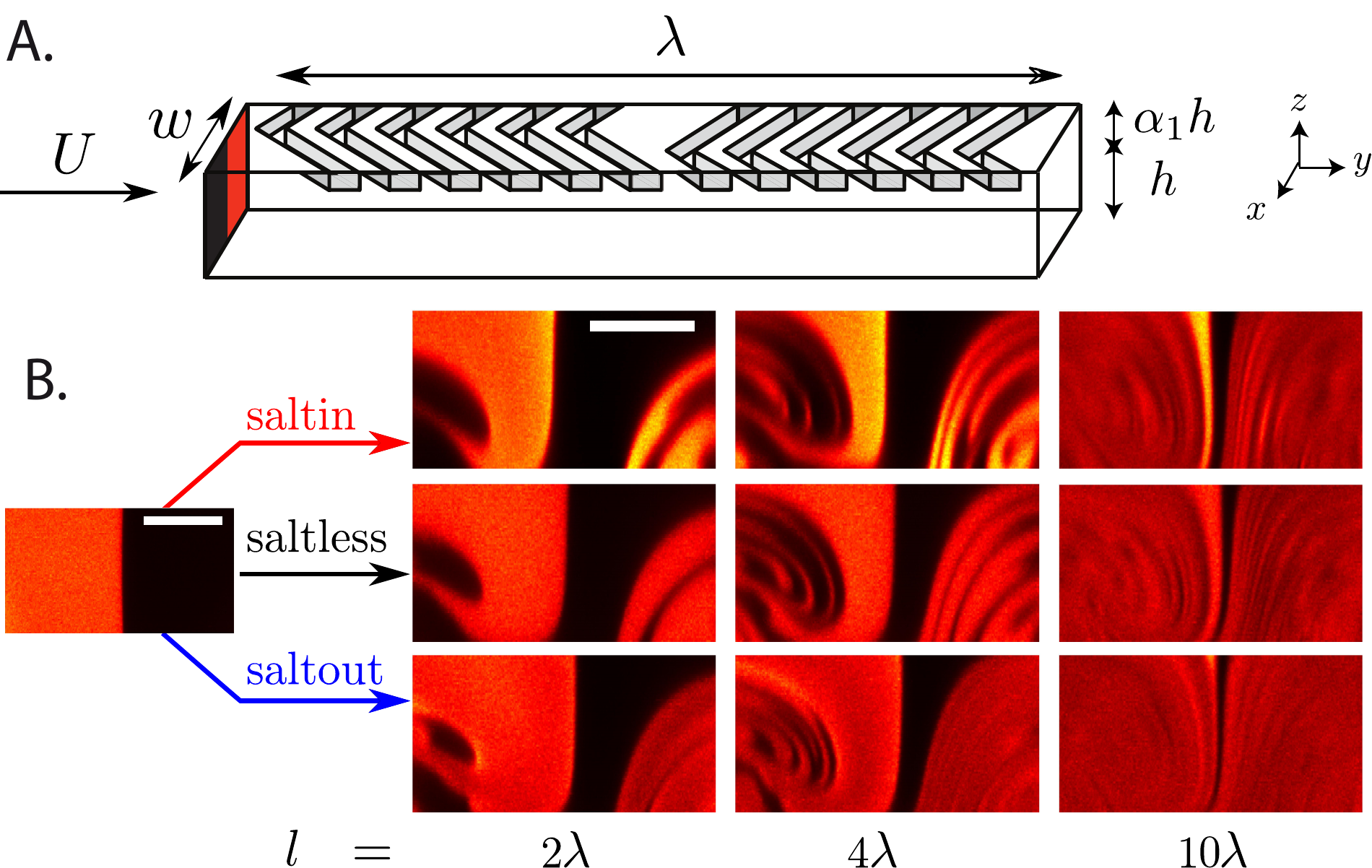}
\caption{Salt effect on chaotic mixing of suspended particles.
\leg{A.} Sketch of chaotic mixing experimental setup: a suspension of fluorescent colloids (orange) and a raw buffer solution (black) are injected at constant flow rate into separate feeding channels (not shown) and merge into the main channel of a Staggered Herringbone Micromixer \cite{stroock2002chaotic} ($w = 200\,\mu\mbox{m}$, $h = 115\,\mu\mbox{m}$, $\lambda = 2\,\mbox{mm}$, $\alpha_1 = 0.35$) where a chaotic mixing of the two solutions takes place.
\leg{B.} Evolution of the suspension mixing process in the presence of salt: cross-section images of the central portion of the mixing channel at different locations $l$ after solutions' merging (white scale bar: $50\,\mu\mbox{m}$; mean flow velocity $U=8.6\,$mm/s). Central row: reference mixing process without salt (\emph{saltless} configuration); Upper and lower rows: effects of additional salt solute ($20\,$mM LiCl) either into colloidal suspension (\emph{salt-in} configuration) or into the co-flowing buffer solution (\emph{salt-out} configuration).
} 
\label{fig:set-up}
\end{figure}

In this communication, we start from this molecular-scale perspective and ask if we can trigger a different molecular transport phenomenon, in place of the canonical diffusion,  and impact in this way the global mixing properties. To explore this bottom-up approach to mixing, we consider an overlooked phenomenon --colloidal \textit{diffusiophoresis}-- recently studied  in the context of surface-driven flow and non-equilibrium transport \cite{prieve1987diffusiophoresis,Munson2002,ajdari2006giant,palacci2010colloidal,palacci2012osmotic}. Local gradients of small chemicals --such as a salt-- induce a migration of particles with transport properties that can be orders of magnitude higher than bare particles diffusion \cite{abecassis2008boosting,abecassis2009osmotic}.
We demonstrate here how this non-equilibrium phenomenon can be harnessed to take control of the molecular-scale transport involved in mixing and thus modify the overall mixing properties of suspended particles.
Indeed we show for the first time that creating an inhomogeneous \textit{chemical} environment by adding only traces of salt deeply alters the mixing dynamics of colloidal suspension, which can be either boosted or inhibited. Moreover, we present a simple model which allows to bridge between the new non-equilibrium nano-scale transport and the large-scale mixing properties.

Experimentally, the mixing of a suspension made of fluorescent colloids (200 nm diameter polystyrene, 0.02\% w/v) with an aqueous buffer solution (1mM Tris, pH=9) is studied under laminar chaotic flows. This is done thanks to a Y-shaped microfluidic device (Fig.~\ref{fig:set-up}A) where the two solutions to mix merge into the main channel of a Staggered Herringbone Micromixer (SHM)  \cite{stroock2002chaotic,stroock2004investigation,aubin2003characterization,yang2005geometric}. Cross-section images are captured with confocal microscope at different channel locations $l$ from the inlet (See Supp. Mat. for further details $^\dag$). This allows to follow the evolution of the mixing process with elapsed time $t=l/U$ or  with the number of stretch and fold cycles $l/\lambda$, with $U$ the mean downstream flow velocity and $\lambda$ the length of one SHM cycle. This SHM was shown to generate a chaotic mixing process \cite{stroock2002chaotic}, see Fig.~\ref{fig:set-up}B middle. 

Now, keeping the global hydrodynamics unchanged, we observe Fig.~\ref{fig:set-up}B that adding a small amount of a \emph{passive} molecular solute ($20\,$mM LiCl salt) to one of the two solutions has a very noticeable impact on the mixing process. Indeed when salt is added to the colloidal suspension (\stin\ configuration, Fig~\ref{fig:set-up}B top), initial stages of mixing show an increased concentration of the colloidal suspension associated with thinner and brighter fluorescent filaments with sharper edges as compared to the \sls\ reference mixing. A ``pinch of salt'' in the particles' phase thus suppresses the homogenization within the mixing process, usually provided by the coupling with local diffusion.

On the contrary, when salt is added to the buffer solution (\sout\ configuration, Fig.~\ref{fig:set-up}B bottom), mixing is again modified but with opposite consequences. Initial stages of mixing now show thicker and dimmer fluorescent filaments with blurred edges (see {\it e.g.} $l/\lambda=4$ in Fig.~\ref{fig:set-up}B). There a trace concentration of salt has a clear enhancing mixing effect. Importantly, we stress that with salt contrasts from $\sim1\,$mM in buffer solvent to $20\,$mM in salty solution, no density or viscosity mismatches nor any floculation or interparticle interaction can be at stake here.


\begin{figure}[!t] 
\centering \includegraphics[width=0.36\textwidth]{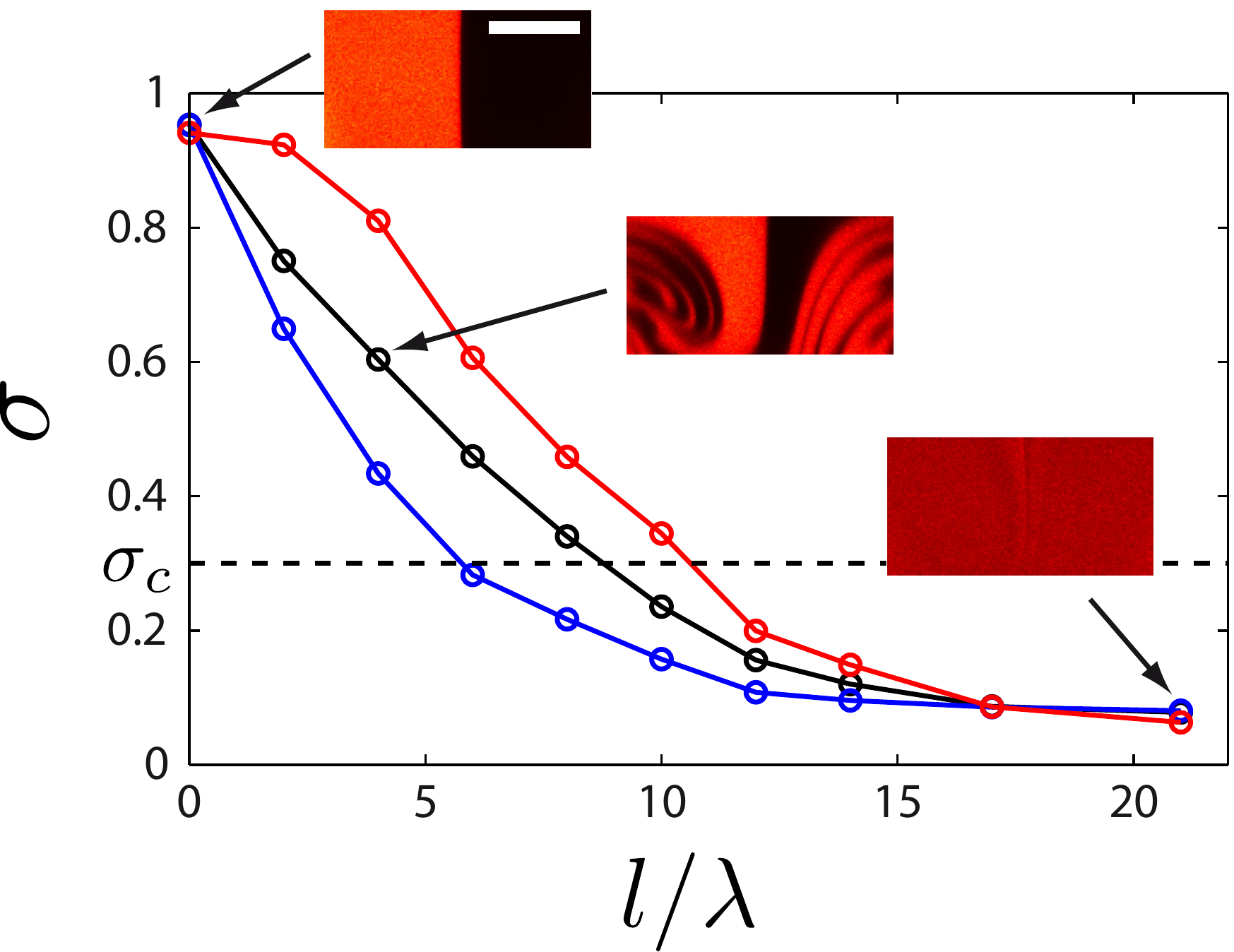}
\caption{Normalized sttandard deviation $\sigma = \sqrt{\langle c ^2 \rangle - \langle c\rangle^2}/\langle c \rangle$ of the colloids concentration $c$  in the channel cross-section as a function of the location $l/\lambda$ for the three mixing configurations (mean flow velocity $U=8.6\,$mm/s): \emph{salt-in} (red), \emph{saltless} (black), \emph{salt-out} (blue). Three corresponding cross-sections are shown for illustration (\emph{saltless} case). Mixing length $l_m$ (see text) is defined at a standard deviation threshold set to $\sigma_c=0.3$.
} 
\label{fig:pictures}
\end{figure}

%
%
%
These observations are confirmed by a more quantitative analysis, using the normalized standard deviation of the colloids concentration field $\sigma = \sqrt{\langle c ^2 \rangle - \langle c\rangle^2}/\langle c \rangle$ to quantify mixing (where the concentration in colloids $c$ was checked to be proportional to the fluorescent intensity). As shown in Fig.~\ref{fig:pictures}, this analysis confirms the previous picture: the presence of salt in the outer solution results in a faster decrease of the concentration inhomogeneities as compared to the \sls\ case, while, on the opposite, salt added to the colloidal suspension delays homogenization. In the following, we define a mixing length $l_m$ using a threshold value as $\sigma(l_m)=\sigma_c=0.3$.

With all flow parameters unchanged, our experiments clearly points to the importance of molecular effects in mixing processes. Such effects are known to occur in the homogenization step, where it is classically assumed that coupling with molecular diffusion allows for local cross-filament transport.
Signature of molecular diffusivity has been reported so far in chaotic  \cite{stroock2002chaotic,stroock2004investigation} or moderately turbulent \cite{villermaux2000mixing,villermaux2006coarse} mixing flows. Qualitatively, mixing is achieved when the diffusional spreading length $L_D$ compares with the typical width of the stretched filaments, say $L_S$. One readily expects $L_D\sim\sqrt{Dl_m/U}$, with $D$ the molecular diffusivity of the mixed species, while in a fully chaotic flow  $L_S\sim(w/2)\exp{(-l_m/\delta)}$, with $w$ the channel width and $\delta$ the stretching length \cite{stroock2002chaotic}. This predicts a weak signature of molecular effects on the mixing length $l_m$ in the form $l_m\sim\log(Uw/D)$.


\begin{figure}[!t] 
\centering \includegraphics[width=0.4\textwidth]{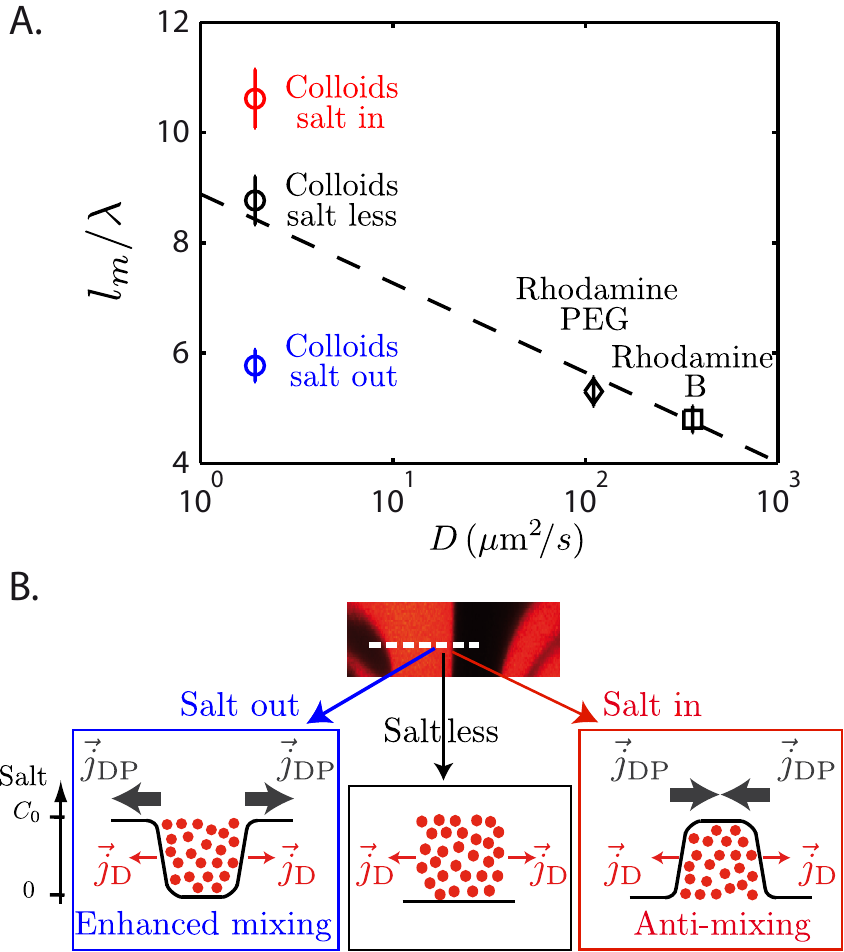}
\caption{\leg{A} Reduced mixing length $l_m/\lambda$ as a function of molecular diffusivity $D$ (mean flow velocity $U=8.6\,$mm/s):  ($\Box$) Rhodamine B dye, ($\Diamond$) Rhodamine-PEG5000, ({\large$\circ$}) 200 nm diameter colloids (saltless, black) and with 20 mM LiCl salt with colloidal suspension (salt-in, red) or with coflowing solution (salt-out, blue).
\leg{B} Diffusiophoretic transport in chaotic mixing: sketch of the underlying salt gradients at the edges of mixing filaments of colloidal suspension. For salt-in and salt-out configurations, total cross-filament flux of colloids (red beads) reads $\vec{j}_c = \vec{j}_D + \vec{j}_{DP}$, with bare diffusive flux $\vec{j}_D$ dominated by orders of magnitude \cite{abecassis2008boosting,palacci2010colloidal} by diffusiophoretic transport $\vec{j}_{DP}$. One thus expects boosted-mixing (salt-out, left) or anti-mixing (salt-in, right).
}
\label{fig:sketchDiffusio} 
\end{figure}

As shown in Fig.~\ref{fig:sketchDiffusio}, this crude argument captures the main effect of molecular diffusivity on mixing  in \sls~ solutions. In this graph, the mixing length $l_m$ is plotted -- for a given flow velocity $U$ -- versus the molecular diffusivity of the suspended particles, from molecules to colloids (see Supp. Mat. for further details $^\dag$), with diffusivities spanning more than 2 orders of magnitude. The measured evolution is compatible with the log dependency expected for chaotic flows, as reported previously for this SHM geometry whereby the effect of changing flow velocity $U$ was probed \cite{stroock2002chaotic}.

Now, as is evident from Fig.~\ref{fig:sketchDiffusio}, the salt effects as a co-mixing solute do not fit into this Brownian diffusion paradigm for molecular transport. 
With colloids diffusivity rigorously unchanged, \stin\ and \sout\ configurations delay or boost mixing as if the molecular transport was changed by about 3 orders of magnitudes (as would be measured in ``diffusivity scale'' fig.~\ref{fig:sketchDiffusio}A). As we show now, this is a striking manifestation of the onset and influence of new out-of-equilibrium local transport (here diffusiophoresis) overtaking classical diffusion.

Diffusiophoretic transport refers to the migration of particles (colloids or macromolecules) induced by gradients of solute. This subtle phenomenon of osmotic origins was studied in pioneering works by Anderson, Prieve, and coworkers \cite{prieve1987diffusiophoresis,anderson1989colloid} and recently received in-depth characterization of its effects on the migration, trapping or patterning of particles \cite{abecassis2008boosting,abecassis2009osmotic,Jiang:2009ke,palacci2010colloidal,palacci2012osmotic}. For salt as a solute with concentration field $C_s$, saline gradients induce a diffusiophoretic drift velocity of a particle as
\begin{equation}
V_{DP} = D_{DP} \nabla\log C_s,
\label{eq:DDP}
\end{equation}
where the diffusiophoretic (DP) mobility $D_{DP}$ has the dimension of a molecular diffusivity \cite{anderson1989colloid}. While the theoretical expression of $D_{DP}$ includes particle's surface charge and salt-type corrections \cite{anderson1989colloid, abecassis2008boosting}, it is typically much larger than the bare colloid diffusion coefficient ($D_{DP}/D\gg1$), and close to small molecules fast diffusivities (here $\DDP \simeq 290\,\mu\mbox{m}^2/s$ \cite{palacci2010colloidal}).

%
%
%

\begin{figure}[!t] 
\centering \includegraphics[width=0.42\textwidth]{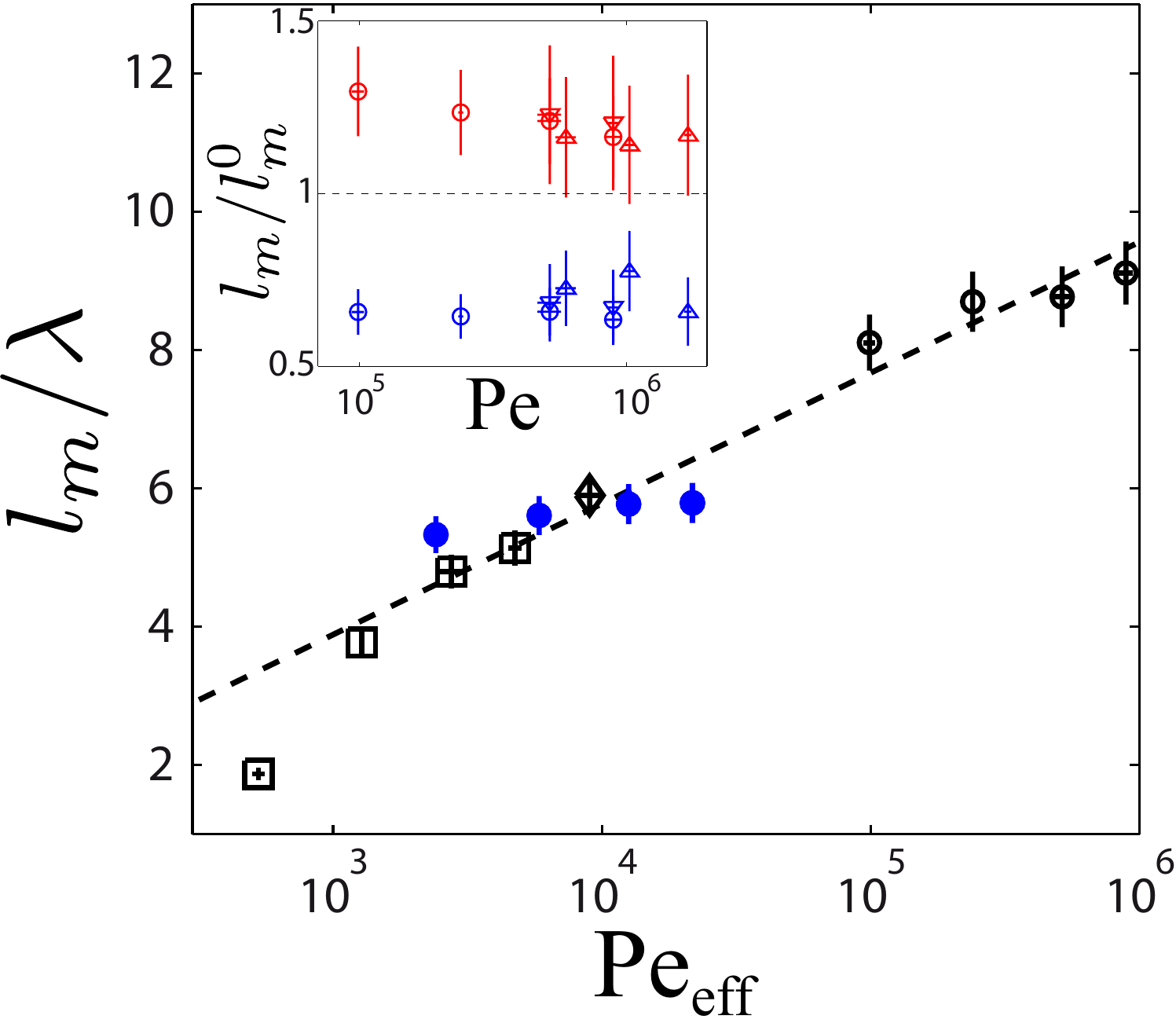}
\caption{Reduced mixing length $l_m/\lambda$ as a function of --effective-- P\'eclet number $\mathrm{Pe}_\mathrm{eff}$ for different velocities, molecular species and salt content (SHM geometrical parameters as in fig.~\ref{fig:set-up}). No salt: rhodamine B ($\Box$), rhodamine PEG5000 ($\Diamond$), \sls\ 200 nm colloids ({\large$\circ$}) ; 20 mM LiCl salt contrast: \sout\ 200 nm colloids ({\large\color{blue}$\bullet$}); $\mathrm{Pe}_\mathrm{eff}$ calculated according to eq. (\ref{eq:lmeff}). The black dashed line represents a $\log$ evolution of $\mathrm{Pe}_\mathrm{eff}$ 
\leg{Inset.} Salt effects on mixing length for colloids $l_m/l_m^0$ as a function of Pe number, for different SHM geometrical characteristics. Salt-less mixing length $l_m^0$ is used as a normalization for \sls\ (red) and \sout\  (blue) data. SHM specifications as in fig.~\ref{fig:set-up} except for $\alpha$: $\alpha_1 = 0.35$ ({\large$\circ$}), $\alpha_2 = 0.40$ ($\bigtriangleup$) and $\alpha_3 = 0.36$ ($\bigtriangledown$). 
}
\label{fig:interplay} 
\end{figure}

Coming back to the colloidal suspension mixing problem, it is now possible to propose a rationalization of the observed salt effects on the basis of this diffusiophoretic mechanism. As sketched in Fig.~\ref{fig:sketchDiffusio}.B, the salt concentration fields $C_s$ exhibits strong gradients localized at the edges of the stretched filaments, thereby triggering a supplementary DP migration of the nearby colloids toward high salt concentrations, see Eq.~\ref{eq:DDP}. This will induce an accelerated spreading of the colloids profile for \sout\, and conversely inhibit mixing for \stin, in full agreement with the observed behavior in Fig.~\ref{fig:sketchDiffusio}.A. This effect is measured systematically for colloids, for all flow velocities or SHM geometries probed as shown in the inset of Fig.~\ref{fig:interplay} where the \stin\ and \sout\ mixing lengths, normalized by the \sls\ reference case $l_m^0$, are plotted against the Peclet number. Finally we need to emphasize that this effect shows up here although the salt concentration remains quite low ($20\,$mM). This is a key consequence of the $\log$ dependency in Eq.~\ref{eq:DDP} allowing response even for traces of solutes as recently demonstrated \cite{palacci2012osmotic}.

Going beyond this analysis towards a more quantitative description of this effect constitutes an important challenge. It involves coupling two intrinsically complex mechanisms: the description of mixing and its complex interplay between flow deformation and local transport; together with the description of molecular-scale non-equilibrium phenomenon controlling here local colloids dynamics.
A first step in this direction can be performed by considering the simplified coupling framework proposed by Ranz for the chaotic laminar mixing process  \cite{ranz1979applications}. It consists in considering the reference frame of a stretched filament, whose axis are aligned with maximal compression and stretching directions, and to reduce the homogenization to a 1D cross-filament dynamics.

For solute species where transport is only ruled by diffusion and flow-advection (\textit{i.e.}  single solute or \sls\ colloids), the Ranz model writes:
\begin{equation}
\partial_t c -\gamma\, x\, \partial_x c = D \partial_x^2 c,
\label{RanzSalt}
\end{equation}
with $x$ the direction perpendicular to the filament interface and $\gamma\propto U$ the principal strain rate. Transverse filament thickness evolves as $s(t)\sim s_0\exp{(-\gamma t)}$, with $s_0=w/2$ the initial width.  This equation is solved by a change of variables $\xi=x/s(t)$ and $\tau=D\int_0^tdt'/s^2(t')$, yielding a simple diffusion equation from which concentration profiles are obtained \cite{ranz1979applications,villermaux2008bridging}. This yields a mixing time for purely advection-diffusion dynamics in the form:
\begin{equation}
t_m^0=\frac{1}{2\gamma}\log\left(\mathrm{Pe}/2 \right),
\label{eq:lm}
\end{equation}
where the Peclet number is defined in the model as $\mathrm{Pe} = \gamma s_0^2/D\propto U w/2D$. This predicts a logarithmic dependence of the mixing length $l_m^0=U\,t_m^0$ in terms of the Peclet number. Despite the simplicity of Ranz model, it is well consistent with our experimental results for single species mixing (Fig.~\ref{fig:sketchDiffusio}.A), in agreement with literature \cite{stroock2002chaotic,stroock2004investigation,villermaux2008bridging}.

It is now possible to extend this framework to include the supplementary colloid DP migration. While the salt concentration $C_s$ is still ruled by Eq.(\ref{RanzSalt}), the cross-filament dynamics of the colloids concentration $c$ in \stin\ and \sout\ now writes:
\begin{equation}
\partial_t c -\gamma\, x\, \partial_x c + \partial_x\left[V_{DP}\, c \right]= D_c \partial_x^2 c,
\label{RanzColl}
\end{equation}
with $V_{DP}$ given by (\ref{eq:DDP}), and $D_c$ the colloids diffusivity. In the reduced variables $(\xi, \tau_s)$ associated to salt dynamics, Eq.(\ref{RanzColl}) transforms to
\begin{equation}
\partial_{\tau_s} c + \partial_\xi\left[\left(\frac{sV_{DP}}{D_s}\right)\, c \right]\simeq0 ,
\label{RanzColl2}
\end{equation}
where a diffusive term of order $D_c/D_s\ll1$ was neglected. Information on the filament interface position $\xi=\Xi(\tau_s)$ can be obtained using the method of characteristics, leading to 
$d\,\Xi/d\tau_s=s(\tau_s)V_{DP}(\Xi(\tau_s),\tau_s)/D_s$.
For short times and small interface displacements $\tau_s, \xi \ll1$, the salt concentration $C_s(\xi,\tau_s)$ reduces to $c_0(1+\xi/\sqrt{\pi\tau_s})/2$ and one may obtain an expression for the interface displacement $\Delta\Xi = \Xi(\tau_s) - \Xi(0)$ as:
\begin{equation}
\Delta\Xi\simeq \pm 2\frac{D_{DP}}{D_s}\sqrt{\frac{\tau_s}{\pi}},
\label{eq:FrontPos}
\end{equation}
valid for short times;  the sign $\pm$ corresponds to the \sout\ and \stin\ cases respectively.

Let us now focus on the  \sout\ case. One defines the --reduced-- mixing time $\tau_s^\mathrm{out}$ as the time when the interface reaches the next colloidal suspension filament ($\xi=\pm1/2$). This is obtained as 
$\tau_s^\mathrm{out} ={D_s}/{2\DF}$,
where we introduced an effective diffusion coefficient defined as $\DF=8\DDP^2/(\pi D_s)$, in line with Ab{\'e}cassis \etal \cite{abecassis2008boosting,abecassis2009osmotic,palacci2010colloidal}. One deduces accordingly, 
\begin{equation}
l_m^\mathrm{out}\equiv U\times t_m^\mathrm{out} =  \frac{U}{2\gamma}\log\left(\mathrm{Pe}_\mathrm{eff}/2\right),
\label{eq:lmeff}
\end{equation}
where we defined an effective P\'eclet number in the form $\mathrm{Pe}_\mathrm{eff}= 2\gamma s_0^2/\DF= \mathrm{Pe} \times (2D_c/\DF)$. 

Comparing with Eq. (\ref{eq:lm}), this suggests that in the \sout\ configuration, the effect of diffusiophoresis can be mapped onto the classical mixing framework, provided the bare molecular diffusivity $D_c$ is replaced by an effective diffusivity $\DF/2$ ($\DF=157\,\mu$m$^2$/s). Accordingly the relevant Peclet number is defined in terms of the {\it effective diffusion} coefficient $\DF$. Since $\DF \gg D_c$, we now expect large particles to mix as fast as small molecules in co-mixing configuration.
While the mixing length is only a weak (logarithmic) function of the Peclet number, the huge change in effective diffusion leads a quantitative change in the mixing efficiency. This prediction is tested in Fig.~\ref{fig:interplay} where the experimental mixing lengths $l_m/\lambda$ obtained for various velocities and salts are plotted against the effective Peclet number $\mathrm{Pe}_\mathrm{eff}$ defined above. 
Although we do not expect this generalized Ranz model to capture the full complexity of the mixing process, it indeed provides an insightful framework that well captures the importance and basic mechanisms of this molecular co-mixing phenomenon.

%
%

Finally, in contrast to the \sout\, configuration, the analysis for the \stin\ case requires accounting for the --up to now neglected-- diffusive transport, which becomes the dominant mechanism for mixing at long time. While the above framework indeed predicts a focusing of the colloid profiles at early times --thus with \emph{no possible diffusivity mapping}--, a full quantitative analysis is far more complex and we leave it for future investigations.


In conclusion, we demonstrate that traces of salt can considerably impact the chaotic mixing of suspensions: here synthetic colloids, but it generalizes to molecules \cite{Munson2002,palacci2010colloidal}, silica \cite{abecassis2008boosting} or clay particles, etc.. This arises from a shift in molecular transport, with canonical diffusion overtaken by diffusiophoresis. Unlike physically-applied fields, the chemical driving force under diffusiophoresis evolves under flow as a mirror of the colloids distribution, thus providing especially relevant and efficient transport unusually bridging across widely separated scales.

Beyond the here-demonstrated interest at microfluidic scale, we stress that diffusiophoresis should show up whenever colloids, particles or molecules evolve in an inhomogeneous chemical background. Combined with the fact that chaotic mixing can be viewed as a prototypal building block for turbulent mixing, this suggests that tuning the local transport properties may yield possible implications at upper-scales \cite{VolkRaynal2014}, in a broad range of situations from bio-reactors to estuaries environment where fresh water full of sediments meets the salty seawater, see {\it e.g.} \cite{thill2001evolution}. Indeed, in the context of marine biology, the local non-equilibrium dynamics of bio-organisms, revealed at the micro-scale, has been hinted as a crucial factor for ocean-scale distribution \cite{paerl1996mini,malits2004effects,stocker2012marine,taylor2012trade,Durham2013}.


\subsubsection*{Acknowledgements}
We thank F. Raynal, M. Bourgoin and R. Volk for helpful discussions, P. Sundararajan for providing initial SHM master and design, G. Simon and R. Fulcrand for technical support. Microfabrication was performed using the NanoLyon facilities (INL). We acknowledge financial support from ANR SYSCOM and from LABEX iMUST (ANR-10-LABX-0064/ ANR-11-IDEX-0007) under project MAXIMIX.





\footnotesize{
\bibliography{mixing_bib} 
\bibliographystyle{rsc} }

\end{document}